\documentclass[prl,twocolumn]{revtex4}
\usepackage{epsfig}
\begin{document}
\flushbottom

\title{Higher order QED in high mass $e^+ e^-$ pairs production
at RHIC}
\author{Anthony J. Baltz$^1$ and Joakim Nystrand$^2$}
\affiliation{$^1$Phyiscs Dept.,
Brookhaven National Laboratory, Upton, NY, 11973, USA\break
$^2$Dept. of Physics and Technology, University of Bergen, Bergen, Norway}

\begin{abstract}
Lowest order and higher order QED calculations have been carried out for
the RHIC high mass $e^+ e^-$ pairs observed by PHENIX with single ZDC triggers.
The lowest order QED results for the experimental acceptance are about
two standard deviations larger than the PHENIX data.
Corresponding higher order QED calculations are within one standard deviation
of the data.
\\
{\bf PACS: { 25.75.-q, 34.90.+q}}
\end{abstract}
\maketitle

A recent publication of the PHENIX collaboration at RHIC has presented data
on ultraperipheral Au + Au photoproduction of the $J/\psi$
and of continuum $e^+ e^-$ pairs in the invariant mass range of
2.0 -- 2.8 GeV/c$^2$\cite{px09}.  The present paper concerns itself
not with the $J/\psi$ but with the continuum pairs.  A previous measurement
of $e^+ e^-$ pairs at RHIC was carried out by the STAR colaboration
in the somewhat lower invariant mass range 140 -- 265 MeV/c$^2$\cite{star04},
and it was subsequently argued that these data exhibited evidence for higher
order QED effects\cite{ajb08}.  This paper presents results of a 
higher order QED calculation for the PHENIX data using the same methodology
as the calculations\cite{ajb08} previously published for the STAR data.

The PHENIX data was presented as a differential cross section with respect to
the pair equivalent mass $m_{e^+e^-}$ and the pair rapidity $y_{pair}$
with the constraint of at least one neutron detected in one of the zero degree
calorimeters (ZDC):
$d^2 \sigma / dm_{e^+e^-} dy_{pair} (e^+ e^- + Xn, y_{pair}=0)$.
The differential cross section at $y_{pair}=0$ was taken from
the sample $\vert y_{pair} \vert < 0.35$
but with no constraint on pseudorapidity $\eta$ of individual electrons and
positrons. However, since the events measured by PHENIX detector had both the
electron and positron pseudorapidities within $\vert \eta \vert < 0.35 $,
the stated results without the individual $\vert \eta \vert < 0.35 $
were necessarily constructed from a model calculation.  The PHENIX publication
states, ``The fraction of events with $\vert(y_{pair})\vert < 0.35$ and
$ 2.0 <m_{e^+e^-} < 2.8$ GeV/c$^2$, where both electron and positron
are within $\vert \eta \vert < 0.35 $ is 1.10\%.  The corresponding
numbers for $2.0 < m_{e^+e^-} <2.3$ GeV/c$^2$ and $2.3 < m_{e^+e^-} <2.8$
GeV/c$^2$ are 1.11\% and 1.08\%, respectively."  Since we wanted
to carry out calculations closest to what PHENIX actually measured, here
we calculate the cross sections with the electron and positron constraints
$\vert \eta \vert < 0.35 $ and compared with the cross sections in the PHENIX
publication multiplied by 0.011, etc. (see also Ref. \cite{ny}).

Following the method of the higher order calculations previously used
for the STAR data,
the cross sections here are computed from the product of the pair production
probability $P_{ee}(b)$, the probability at least one of the ions Coulomb
dissociating  $P_{1x}(b)$, and a survival factor $\exp[-P_{nn}(b)]$ to exclude
events where hadronic interactions occur
\begin{equation}
\sigma = 2 \pi \int P_{1x}(b) P_{ee}(b) \exp[-P_{nn}(b)] b db .
\end{equation}
Unlike the STAR Coulomb dissociation factor $P_{xx}(b)$, which corresponds to
neutrons detected in both ZDCs,
\begin{equation}
P_{xx}(b) = [1 - \exp(-P_C(b)]^2 ,
\end{equation}
here the corresponding Coulomb dissociation factor
$P_{1x}(b)$ is the unitarized probability that requires only that at least one
of the colliding nuclei suffers Coulomb dissociation
\begin{equation}
P_{1x}(b) = [1 - \exp(-2 P_C(b)] .
\end{equation}
$P_C(b)$ the non-unitarized probability of a single Coulomb excitation
calculated in a phenomenological
model for the ZDC neutrons derived from photodissociation
data\cite{brw,bcw}.
The non-unitarized hadronic interaction probability $P_{nn}(b)$ is calculated
in the usual Glauber manner
\begin{equation}
P_{nn}(b) = \sigma_{nn}\int dx dy \ T_A(x,y)\  T_A(x-b,y)].
\end{equation}
where $\sigma_{nn}$ is the total hadronic interaction cross section, 52 mb
at RHIC, $b$ is the impact paramete, and the nuclear thickness function
$T_A(x',y)$ is the longitudinal integral of the nuclear density, $\rho(r)$
\begin{equation}
T_A(x',y) = \int dz \rho(x',y,z) dz.
\end{equation} 
The nuclear density profile $\rho(r=\sqrt{x'^2+y^2+z^2})$ of heavy
nuclei is well described with a Woods-Saxon distribution.  We use
parameters determined from electron scattering data (R=6.38 fm a=.535 for Au).

The perturbative and higher order QED pair production probabilities $P_{ee}(b)$
were calculated using the methods previously described\cite{ajb06,ajb08}.
In the $P_{ee}(b)$ calculations
an analytical elastic form factor was employed
\begin{equation}
f(q)={3 \over (q r)^3} [ \sin(q r) - q r \cos(q r) ]
\Biggl[{ 1 \over 1 + a^2 q^2} \Biggr]
\end{equation}
with a hard sphere radius $r = 6.5$ fm and Yukawa potential range
$a = 0.7$ fm.  This form very closely reproduces the Fourier transformation
of the Au density with the Woods-Saxon parameters 
mentioned above\cite{kn}.

Calculations in Ref.\cite{px09} performed with
the computer program Starlight\cite{bgkn} differed from the present
calculation only in the use of an equivalent photon probability
$P^{ww}_{ee}(b)$ in Eq. (1) instead of the QED probability $P_{ee}(b)$.
The Coulomb dissociation probability $P_{1x}(b)$ and hadronic survival
probabilities $\exp[-P_{nn}(b)]$ were treated as described
above identically in both calculations.
The equivalent photon probability $P^{ww}_{ee}(b)$ uses the Weizsacker-Williams
photon spectrum calculated for the heavy ions and the Breit-Wheeler cross
section for $\sigma(\gamma + \gamma \rightarrow e^+e^-)$\cite{bgkn}.
In Starlight it is also required that photons originate only outside
the nuclear radius $R$ of each of the ions, and in the
PHENIX calculations, a hard sphere radius $R=6.98$ fm was used for this
sharp cutoff.  

\begin{table}
\caption[Table I]{RHIC: Au + Au, $\gamma = 107$, single ZDC triggered
$e^+ e^-$ pairs cross section ($d\sigma/dm_{e^+e^-}dy_{pair} \; [\mu b/GeV]$) 
with PHENIX cuts (see text).}
\begin{tabular}{|c|cccc|}
\colrule
 $m_{e^+e^-}$ & Data & Starlight & Perturbative & Higher Order \\
 (GeV/c$^2$) &\cite{px09,ny} & \cite{bgkn} & QED \cite{ajb08,ajb06} & QED \cite{ajb08,ajb06} \\
\colrule
2.0 -- 2.8   & 0.95 $\pm$ 0.31 & 0.99 & 1.69 & 1.24 \\
2.0 -- 2.3   & 1.43 $\pm$ 0.61 & 1.53 & 2.60 & 1.90 \\
2.3 -- 2.8   & 0.65 $\pm$ 0.30 & 0.66 & 1.14 & 0.83 \\
\colrule
\end{tabular}
\label{tabi}
\end{table}

Results are shown in Table I.  The stated statistical and systematic errors
in the data have been combined in quadrature.
As indicated above, calculations are for the measured Au + Au reaction
at $\gamma = 107$ for each of the
Au nuclei.  $e^+e^-$ pairs were accepted for $\vert y_{pair} \vert < 0.35$
with a ZDC trigger indicating simultaneous dissociation of at least one
of the gold nuclei.
Individual electrons and positrons were also accepted only for mid-rapidity
$\vert \eta \vert < 0.35$.  The calculated cross section is defined for
a pair to literally fall within the cuts: $y_{pair}$, individual $e^+$
and $e^-$ $\vert \eta \vert$, and $m_{e^+e^-}$,
but normalized to unit $y_{pair}$ and $m_{e^+e^-}$.

As was shown in the original PHENIX publication, the Starlight calculations
are in very good agreement with the data.
The present QED perturbative calculations are about two standard deviations
higher than the data, while the higher order
calculations are higher than the data but within one standard deviation for 
all three $m_{e^+e^-}$ ranges.

One might reasonably ask why the peturbative QED calculations are so much
higher than the perturbative Weizsacker-Williams calculations.  In the
first place one has to recognize that Weizsacker-Williams calculations are
an appproximation to the more proper QED calculations, in particlar neglecting
the viruality of the intermediate photons.  However there is another
physics difference between the two calculations.
In the QED calculation there is no sharp cutoff analagous to the  
cutoff at the hard sphere nuclear radii $R$ 
in the STAR equivalent photon calculation, but the
form factor, Eq. (6) gives a smooth cutoff in both tranverse and
longitudital photon  momenta arising from the nuclear
Woods-Saxon charge distribution.  This smooth cutoff allows more
high momentum photons contributing to the cross section
than a sharp cutoff does and should increase the cross section.

A simple exploratory luninosity calculation can be carried out for the case
of a uniform hard sphere nuclear charge distribution by allowing contributions
from within the ions but with an effective Z multiplied by $(b_i/R)^3$, where
$b_i$ is the distance of the photon source from the center of the nucleus.
The result is that with this
finite density cutoff the two photon luminosity in the STAR measured equivalent
pair mass range would increase by 44\% over the sharp cutoff,
and one can infer a proportional increase
in the predicted perturbative Weizsacker-Williams pair cross sections.

The physical question is whether equivalent photons arising from within
one the ions when the ions are not overlaping should contribute to the process.
We have previously noted\cite{bgkn} that the requirement that final
states be outside the nuclei may not be necessary for some final states
such as lepton pairs.  The smooth cutoff conventionally implemented by a
momentum dependent form factor in QED calculations does not seem unreasonable.

One of us (AJB) would like to acknowledge useful discussions with Sebastian
White and Mickey Chiu.

This manuscript has been authored
under Contract No. DE-AC02-98CH10886 with the U. S. Department of Energy.

\end{document}